\documentclass[aps,prl,nopacs,nokeys,superscriptaddress,10pt,twocolumn,amsfonts,asmsymb]{revtex4}

\usepackage{graphicx,epic,eepic,epsfig,amsmath,latexsym,amssymb,verbatim,revsymb}

\usepackage{color}
\usepackage{theorem}

\def\squareforqed{\hbox{\rlap{$\sqcap$}$\sqcup$}}
\def\qed{\ifmmode\squareforqed\else{\unskip\nobreak\hfil
\penalty50\hskip1em\null\nobreak\hfil\squareforqed
\parfillskip=0pt\finalhyphendemerits=0\endgraf}\fi}
\def\endenv{\ifmmode\;\else{\unskip\nobreak\hfil
\penalty50\hskip1em\null\nobreak\hfil\;
\parfillskip=0pt\finalhyphendemerits=0\endgraf}\fi}

\mathchardef\ordinarycolon\mathcode`\:
\mathcode`\:=\string"8000
\def\vcentcolon{\mathrel{\mathop\ordinarycolon}}
\begingroup \catcode`\:=\active
  \lowercase{\endgroup
  \let :\vcentcolon
  }

\newcommand{\nc}{\newcommand}
\nc{\rnc}{\renewcommand}
\nc{\beq}{\begin{equation}}
\nc{\eeq}{{\end{equation}}}
\nc{\beqa}{\begin{eqnarray}}
\nc{\eeqa}{\end{eqnarray}}
\nc{\lbar}[1]{\overline{#1}}
\nc{\bra}[1]{\langle#1|}
\nc{\ket}[1]{|#1\rangle}
\nc{\ketbra}[2]{|#1\rangle\!\langle#2|}
\nc{\braket}[2]{\langle#1|#2\rangle}
\nc{\proj}[1]{| #1\rangle\!\langle #1 |}
\nc{\avg}[1]{\langle#1\rangle}
\rnc{\max}{\operatorname{max}}
\nc{\Rank}{\operatorname{Rank}}
\nc{\smfrac}[2]{\mbox{$\frac{#1}{#2}$}}
\nc{\tr}{\operatorname{Tr}}
\nc{\ox}{\otimes}
\nc{\dg}{\dagger}
\nc{\dn}{\downarrow}
\nc{\cA}{{\cal A}}
\nc{\cB}{{\cal B}}
\nc{\cC}{{\cal C}}
\nc{\cD}{{\cal D}}
\nc{\cE}{{\cal E}}
\nc{\cF}{{\cal F}}
\nc{\cG}{{\cal G}}
\nc{\cH}{{\cal H}}
\nc{\cI}{{\cal I}}
\nc{\cJ}{{\cal J}}
\nc{\cK}{{\cal K}}
\nc{\cL}{{\cal L}}
\nc{\cM}{{\cal M}}
\nc{\cN}{{\cal N}}
\nc{\cO}{{\cal O}}
\nc{\cP}{{\cal P}}
\nc{\cR}{{\cal R}}
\nc{\cS}{{\cal S}}
\nc{\cT}{{\cal T}}
\nc{\cX}{{\cal X}}
\nc{\cZ}{{\cal Z}}
\nc{\csupp}{{\operatorname{csupp}}}
\nc{\qsupp}{{\operatorname{qsupp}}}
\nc{\var}{\operatorname{var}}
\nc{\rar}{\rightarrow}
\nc{\lrar}{\longrightarrow}
\nc{\polylog}{\operatorname{polylog}}
\nc{\1}{{\openone}}

\nc{\RR}{{{\mathbb R}}}
\nc{\CC}{{{\mathbb C}}}
\nc{\FF}{{{\mathbb F}}}
\nc{\NN}{{{\mathbb N}}}
\nc{\ZZ}{{{\mathbb Z}}}
\nc{\PP}{{{\mathbb P}}}
\nc{\QQ}{{{\mathbb Q}}}
\nc{\UU}{{{\mathbb U}}}
\nc{\EE}{{{\mathbb E}}}
\nc{\id}{{\operatorname{id}}}

\nc{\be}{\begin{equation}}
\nc{\ee}{{\end{equation}}}
\nc{\bea}{\begin{eqnarray}}
\nc{\eea}{\end{eqnarray}}
\nc{\<}{\langle}
\rnc{\>}{\rangle}
\nc{\Hom}[2]{\mbox{Hom}(\CC^{#1},\CC^{#2})}
\nc{\rU}{\mbox{U}}

\nc{\ob}[1]{#1}

\begin{document}

%\title{The $\mathbf{W + \overline{W}}$ state is genuinely multiparty entangled}

%\title{Quantum Correlation Without Classical Correlations?}

\title{Quantum Correlation Without Classical Correlations}

\author{Dagomir Kaszlikowski}
\affiliation{Quantum Information Technology Lab, National University of Singapore,
 2 Science Drive 3, 117542 Singapore, Singapore}
\author{Aditi Sen(De)}
\affiliation{ICFO-Institut de Ci\`encies Fot\`oniques,
Mediterranean Technology Park,
E-08860 Castelldefels (Barcelona), Spain}
\author{Ujjwal Sen}
\affiliation{ICFO-Institut de Ci\`encies Fot\`oniques,
Mediterranean Technology Park,
E-08860 Castelldefels (Barcelona), Spain}
\author{Vlatko Vedral}
\affiliation{The School of Physics and Astronomy, University
of Leeds, Leeds, LS2 9JT, United Kingdom}
\affiliation{Quantum Information Technology Lab, National University of Singapore,
 2 Science Drive 3, 117542 Singapore, Singapore}
\author{Andreas Winter}
\affiliation{Department of Mathematics, University of Bristol, Bristol BS8 1TW, U.K.}
\affiliation{Quantum Information Technology Lab, National University of Singapore,
 2 Science Drive 3, 117542 Singapore, Singapore}

%\date{10 April 2007}
%date{Barca thhekey 16 April 2007}
%date{Barca thhekey abar 19 April 2007}

%\date{13 May 2007}

\begin{abstract}

We show that genuine multiparty quantum correlations can exist on its own, without a supporting 
background of genuine multiparty classical correlations,
even in macroscopic systems. Such possibilities can have important implications in the physics of quantum information and phase transitions.

\end{abstract}

\maketitle

%Quantum and classical correlations lie at the heart of sciences and technologies.
%The importance of correlations in sciences and mathematics and beyond can hardly be overemphasized. 
%Quantum physical laws describe correlations between preparations and measurement
%outcomes of systems under study. 
%In biology, DNA replication relies on correlations between the four bases making up the DNA. 
%The field of statistics in mathematics has grown out of our need for a deeper description of correlated events in nature. 
%But the importance of correlations stretches far beyond sciences and mathematics. 

Quantum and classical correlations lie at the heart of sciences and technologies.
%Quantum physical laws describe correlations between preparations and measurement
%outcomes of systems under study. 
%In biology, DNA replication relies on correlations between the four bases making up the DNA. 
%The field of statistics in mathematics has grown out of our need for a deeper description of correlated events in nature.
%
The emerging quantum technology 
crucially depends on correlations that are different and more subtle 
than the ones in
classical physics. 
This quantum form of correlations, known as entanglement \cite{HORO-RMP}, is currently being exploited to achieve  
higher levels of security in cryptography \cite{GISIN-RMP} and  faster rates of information 
processing in computers \cite{nielsenchuang}.
Since quantum physics contains the classical one as a special case, we 
may think that reducing entanglement will 
in the same way ultimately lead to the reduction of all correlations 
to classical correlations. 
We will however show that this intuition is completely false; surprisingly, 
genuine multiparty entanglement can exist on its own without the need 
of genuine multiparty classical correlations, even for macroscopic systems. 
This result has potentially fundamental implications in the physics of quantum information and phase transitions, 
and in the current characterization of the quantum to classical transition.

It is widely known that quantum correlations are of a different
kind than classical ones, and that quantum correlations
can be vanishing even when classical correlations are present.
However, the fact that even the opposite is true has so far eluded us. 
We present a general method to 
obtain that multiparty quantum and classical correlations 
are independent for a certain class of multiparty states, shared between an arbitrary odd number of two-dimensional systems (qubits). 
This phenomenon cannot happen for quantum 
systems with two subsystems; neither can it happen for pure
multiparty states.

{\bf Quantum and classical correlations.}
We begin by discussing in generality how to distinguish quantum correlations
and specifically genuine multiparty correlations, as well as the
analogous classical notions.
For an $n$-party state $\rho$, it is established usage to call it
separable (or, precisely, \emph{fully} separable) if it is a probability
mixture (i.e., convex combination) of $n$-party product states \cite{Werner891}; otherwise
the state is (somehow) entangled.
A more stringent notion is that of genuine $n$-party entanglement~\cite{genuine},
which demands that the state is not in the convex hull of tensor
products over any bipartition of the $n$ systems. [See~\cite{genuine} for
a whole hierarchy of notions, $k$-separability and the complement,
genuine $(n+2-k)$-party entanglement, for $k=2,\ldots,n$.]

All this is fairly canonical -- though one might consider the idea of
defining quantum correlations via the violation of some Bell
inequality \cite{EPRBell2}, which is a strictly different concept, already in the
bipartite case~\cite{Werner891, Werner892}. We shall come back to this issue,
but for the present paper, we stick with this more encompassing notion.
On the other hand, defining correlations in general, and in particular
genuinely multiparty \emph{classical} correlation in a state, seems
more contentious.

We propose the following point of view. First of all, classical
correlations are to be about the values of (local) observables.
Furthermore, taking the bipartite case as a model, it is easy to
see that a state $\rho_{AB}$ is correlated (i.e., not equal to a product
state $\rho_A \otimes \rho_B$) if and only if there are local
observables $X$ and $Y$ such that the \emph{classical} variables
$X$ and $Y$ are correlated. This in turn can be determined by
looking at covariances $\text{Cov}(X,Y) = \bigl\< (X-\<X\>)(Y-\<Y\>) \bigr\>$:
it is again not hard to see that if these are zero for all choices
of local observable then the state must be a product, and vice versa.
Unless mentioned otherwise,  we shall restrict here to traceless observables $X$, $Y$, and if the
states under consideration are such that $\<X\> = \<Y\> = 0$,
then the covariances mentioned reduce to correlators
$\<XY\> = \tr\rho(X\otimes Y)$, well-known quantities in statistical
physics.

Encouraged by these observations, we take as our (at least
sufficient) criterion of genuine \(n\)-party
%Encouraged by these observations, we will take as our notion
%of genuine $n$-party 
correlation, in a state $\rho_{12\ldots n}$,
that for some choice of local observables $X_j$, the ``covariance''
$\text{Cov}(X_1,\ldots,X_n) = \bigl\< (X_1-\<X_1\>)\cdots(X_n-\<X_n\>) \bigr\>$
is nonzero. 
%%%%%%%%%%%%%%%%
The question
how to \emph{define} genuine \(n\)-party correlations has been
considered before; e.g. Zhou \emph{et al.} \cite{Zhouetal} have proposed
a set of axioms for \(n\)-party correlation measures. It can
be shown that, at least for three parties and states with
maximally mixed marginals, our criterion implies that of
Ref. \cite{Zhouetal}.
%%%%%%%%%%%%%%%
We shall only have occasion to look at states with
maximally mixed marginals, so that it is enough if we stick to traceless observables, in which
case this expression reduces to the higher correlator
$\<X_1\cdots X_n\> = \tr\rho(X_1\otimes\cdots\otimes X_n)$.
It is in this sense that we shall examine the presence of
genuine $n$-party classical correlations in a state.
We do not claim to have a well-established definition of genuine n-party correlations. 
All we want to say is that if all n-party covariances of an n-party state vanish, 
then the state should be considered to have no genuine n-party correlations.

It is curious to remark that the majority of Bell's inequalities,
including multiparty ones, is actually expressed in terms of 
$n$-party correlators \cite{zukowski1, zukowski2, zukowski3, zukowski4}.
Also, note that for an $n$-qubit state, as we shall look at, the absence
of genuine $n$-party (more generally $k$-party) correlations means
just that in the Pauli basis expansion all terms of $n$ ($k$)
proper Pauli ($\{\sigma_x,\sigma_y,\sigma_z\}$) operators are vanishing.

{\bf The example.}
For an integer $n \geq 3$, consider the $W$-state \cite{W1,W2} of \(n\) 
two-dimensional quantum systems (qubits)
\(
  \ket{W} = \frac{1}{\sqrt{n}} \bigl( \ket{00\ldots 001}
                                     +\ket{00\ldots 010}
                                     +\ldots
                                     +\ket{10\ldots 000} \bigr),
\)
where \(|0\rangle\) and \(|1\rangle\) are eigenstates of the \(\sigma_z\) Pauli operator, 
and its ``complement''
%\[\begin{split}
 \( \ket{\overline{W}} = \frac{1}{\sqrt{n}} \bigl( \ket{11\ldots 110}
                                                 +\ket{11\ldots 101}
                                                 +\ldots
                                                 +\ket{01\ldots 111} \bigr) 
                     = \sigma_x^{\ox n}\ket{W},
\)
%\end{split}\]
%
and form the state
\[
  \rho_p = p \proj{W} + (1-p)\proj{\overline{W}},
\]
which is just a mixture of \(\ket{W}\) and \(\ket{\overline{W}}\) with
probabilities \(p\) and \(1-p\). 
We must of course have \(0\leq p \leq 1\).

{\bf Quantum correlations exist.} We first show that this state is genuinely multiparty entangled;
in fact, we show that the subspace ${\cal S}$ spanned by $\ket{W}$ and
$\ket{\overline{W}}$ contains no product vector $\ket{\varphi}_A \ket{\psi}_B$
for any partition $A \cup B$ of the sites $\{1,\ldots,n\}$.
The subspace \({\cal S}\) is the support of the state \(\rho_p\): Any decomposition of 
\(\rho_p\) must be a probabilistic mixture of pure states from \({\cal S}\). Therefore if we are able to show that 
\({\cal S}\) does not contain any product state, \(\rho_p\) cannot be written as a probabilistic mixture of product pure states 
in any bipartite splitting. Consequently, \(\rho_p\) will be genuinely multiparty entangled.

The subspace \({\cal S}\)
%under consideration 
has a lot of symmetry,
which we will exploit: it is invariant under permutations 
of the sites, and under bit flip at all the sites (i.e. under $\sigma_x^{\ox n}$).
On the other hand, it is only
two-dimensional, which will eventually result in a contradiction
if we assume the existence of a product unit vector
$\ket{\varphi}_A \ket{\psi}_B \in {\cal S}$.

By the permutation symmetry, we may assume, without loss of generality, that
$A=\{1,\ldots,k\}$ and $B=\{k+1,\ldots,n\}$. Now, we first focus on
the ``weights'' of the vectors involved, i.e., the numbers of $\ket{1}$'s
in the expansion in the computational basis (the \(\sigma_z^{\ox n}\) eigenbasis).
Observe that in ${\cal S}$,
only weights $1$ and $n-1$ occur. Let the set of weights occurring in
$\ket{\varphi}$ be $F$, and $G$ the corresponding set for $\ket{\psi}$;
then the weights in the tensor product $\ket{\varphi}_A \ket{\psi}_B$
are precisely all pairwise sums, $F+G$. Since this has to be a subset
of $\{1,n-1\}$, only $|F|=1$, $|G|\leq 2$ and symmetrically $|G|=1$,
$|F|\leq 2$ are possible, and we may, without loss of generality, restrict to the former
possibility.  (\(|\Lambda|\) denotes the cardinality of the set \(\Lambda\).) 
Therefore only one weight, say $r$, occurs in $\ket{\varphi}$.
%This means that in $\ket{\varphi}$ only one weight
%occurs, say $r$.

Now consider the projectors $P_1$  and $P_{n-1}$ in the $n$-qubit system, where
$P_k$ projects onto the subspace spanned by basis vectors of weight $k$.
Clearly, $P_1 {\cal S}$ is just the line spanned by $\ket{W}$,
and $P_{n-1} {\cal S}$ is 
that 
%the line spanned 
by $\ket{\overline{W}}$.

The first observation is that
\begin{align}
  \label{eq:P1}
  P_1     \ket{\varphi}_A \ket{\psi}_B &= \ket{\varphi}_A \ket{\psi'}_B, \\
  \label{eq:Pn-1}
  P_{n-1} \ket{\varphi}_A \ket{\psi}_B &= \ket{\varphi}_A \ket{\psi''}_B,
\end{align}
with vectors $\ket{\psi'}$ and $\ket{\psi''}$, which have only one weight 
occurring, namely $1-r$ and $n-1-r$, respectively. This is because
$\ket{\varphi}$ is already in a constant weight subspace, so the projection
only ends up affecting $\ket{\psi}$.

But we observed already that the vectors in Eqs.~(\ref{eq:P1}) and
(\ref{eq:Pn-1}) must be proportional to $\ket{W}$ and $\ket{\overline{W}}$,
respectively, which are not product across any cut, so the right
hand sides above must both be zero, and we arrive at the desired
contradiction.
This concludes the proof of the statement that the state \(\rho_p\) has genuine multiparty entanglement,
whenever the state is composed of three or more qubits, and for any \(p\) in \([0,1]\).

Before proceeding further, let us note that the key ingredients in the above demonstration were the symmetries of the state \(\rho_p\),
and the low dimensionality of the support \({\cal S}\) of \(\rho_p\). We expect to be able to relax the second 
ingredient -- other small, but
more than two-dimensional subspaces will have the same property
of being ``completely entangled''.
The first property should also not be strictly necessary, since it is clear
that with ${\cal S}$, also a sufficiently small perturbation ${\cal S}'$
will contain no product vectors.

{\bf No classical correlations.} We now show that for any \emph{odd} number \(n\) (greater than one) of 
qubits, the state \( \rho := \rho_{1/2}\) 
does \emph{not} have any genuine \(n\)-partite classical correlations. 
More precisely, we show that the average value of \emph{any} tensor
product of traceless observables in the state is vanishing.
%Let us find the \(n\)-partite classical correlation tensor \cite{W3}
%\(t_{A_1  A_2 \ldots A_n}^W = \langle W| \sigma_{A_1} \ox \sigma_{A_2} \ldots \ox \sigma_{A_n} |W\rangle \),
%where \(\sigma_{A_i}\) (\(i=1,2, \ldots, n\)) are chosen from among the Pauli operators  $\sigma_x$, $\sigma_y$ and 
%$\sigma_z$,
%for the \(W\) state. 
%We have  
%\(t_{zz \ldots z}^W = - 1\), and \(t_{A_1  A_2 \ldots A_n}^W =2/n\), 
%whenever exactly two the \(\sigma_{A_i}\) are the flipping operators
%\(\sigma_x\) or \(\sigma_y\), and the rest are all \(\sigma_z\). This is 
%true for both even and odd \(n\). 
%In contrast, for odd \(n\),  the \(n\)-partite classical correlation tensor
%\(t_{A_1  A_2 \ldots A_n}^{\overline{W}}\) of the \(\overline{W}\) state 
%is  just \(-t_{A_1  A_2 \ldots A_n}^W\),
%while for even \(n\), 
%\(t_{A_1  A_2 \ldots A_n}^{\overline{W}} = t_{A_1  A_2 \ldots A_n}^W\).
%This is a consequence of the $\sigma_z^{\otimes n}$ symmetry of $\rho$.
%
%
 First, we observe that $\rho$
%_N$ ($N$ is the number of qubits) 
can be written as an equal mixture of the states
 %\begin{equation}
 \(|V_{\pm}\rangle = \frac{1}{\sqrt{2}}(|W\rangle\pm|\overline{W}\rangle)\).
%\end{equation}
%
We have 
%
%\begin{equation}
\(\sigma_x^{\otimes n}|V_{\pm}\rangle = \pm |V_{\pm}\rangle\).
%\end{equation}
%
Thus
\begin{eqnarray}
&&\langle V_{\pm}|\sigma_{k_1}^{A_1}\otimes\dots\otimes\sigma_{k_n}^{A_n}|V_{\pm}\rangle 
= \langle V_{\pm}|\bar{\sigma}_{k_1}^{A_1}\otimes\dots\otimes\bar{\sigma}_{k_n}^{A_n}|V_{\pm}\rangle\nonumber\\
&&=(-1)^{n_2+n_3}\langle V_{\pm}|\sigma_{k_1}^{A_1}\otimes\dots\otimes\sigma_{k_n}^{A_n}|V_{\pm}\rangle,
\end{eqnarray}
where \(n_1\), $n_2$, and $n_3$ are respectively the numbers of occurences 
of the Pauli matrices $\sigma_x$, $\sigma_y$,  and $\sigma_z$,
 in the sequence $\sigma_{k_1}^{A_1}\otimes\dots\otimes\sigma_{k_n}^{A_n}$, with \(\sigma_{k_i}^{A_i}\) 
being a Pauli operator at the \(i\)th site (\(i=1,2, \ldots, n\), \(k_i =x,y,z\)), and 
\(\bar{\sigma}_{k_i} = \sigma_x\sigma_{k_i}\sigma_x\). It is clear 
that if $n_2+n_3$ is an odd number, we have 
$\mbox{Tr}(\rho\sigma_{k_1}^{A_1}\otimes\dots\otimes\sigma_{k_n}^{A_n})=0$. If $n_2+n_3$ is an even number, we have 
\begin{equation}
\mbox{Tr}(\rho\sigma_{k_1}^{A_1}\otimes\dots\otimes\sigma_{k_n}^{A_n})=
\langle W| \sigma_{k_1}^{A_1}\otimes\dots\otimes\sigma_{k_n}^{A_n}|W\rangle.
\end{equation}
Due to the permutation symmetry of the state $|W\rangle$, the above expression is equal to 
\begin{equation}
\langle W|\bigotimes_{i=1}^{n_1}\sigma_x^{A_i}\bigotimes_{i=n_1 +1}^{n_1+n_2}\sigma_y^{A_i}
\bigotimes_{i=n_1+n_2+1}^{n}\sigma_z^{A_i}|W\rangle,
\label{expr}
\end{equation}
with, as above, $n_2+n_3$ being an even number. For convenience, let us define 
$\Sigma(n_1,n_2,n_3) = \bigotimes_{i=1}^{n_1}\sigma_x^{A_i}\bigotimes_{i=n_1 +1}^{n_1+n_2}\sigma_y^{A_i}
\bigotimes_{i=n_1+n_2+1}^{n}\sigma_z^{A_i}$.
%
%\bigotimes_{k_1=1}^{n_1}\sigma_1^{(k_1)}\bigotimes_{k_2=1}^{n_2}\sigma_2^{(k_2)}
%\bigotimes_{k_3=1}^{n_3}\sigma_3^{(k_3)}$. 
The state $|W\rangle$ is a superposition of $n$ pure \(n\)-qubit states having 
only one qubit in the state $|1\rangle$ and the rest
in the state $|0\rangle$. This means that \(\Sigma(n_1,n_2,n_3)|W\rangle \)
\begin{eqnarray}
&& = i\delta_{n_1+n_2,2}\frac{(-1)^{n-2}}{\sqrt{n}}(|100\dots0\rangle-|010\dots 0\rangle)+\nonumber\\
&&(1-\delta_{n_1+n_2,2})|W(n_1,n_2,n_3)\rangle,
\end{eqnarray}
where $\langle W(n_1,n_2,n_3)|W\rangle = 0$. Clearly, 
%the above formula 
this
implies that $\langle W|\Sigma(n_1,n_2,n_3)|W\rangle=0$, which ends the proof.

{\bf Discussion.}
Consequently, for odd \(n\), the state 
\[
  \rho = \frac{1}{2}|W\rangle \langle W| + \frac{1}{2}|\overline{W}\rangle \langle \overline{W}|
\]
has no genuine \(n\)-partite classical correlations, despite the fact, as we have already shown, that 
it has genuine \(n\)-partite quantum correlations. Note, as an aside, that it
\emph{does} have $(n-1)$-party correlations, in the sense specified initially.
Note also that if instead of looking at correlators, we look at the covariances 
\(\text{Cov}(X_1{'},\ldots,X_n{'})\), of arbitrary observables \(X_1{'},\ldots,X_n{'}\) (which are not 
necessarily traceless), then such covariances vanish for
% the state 
\(\rho\).

The states of rank unity (pure states) do not lend themselves to the phenomenon under study, while already 
rank-two states are shown to be eligible. Also, while bipartite systems
cannot show this phenomenon, already a three-qubit 
system is qualified; for three qubits, an example of a state like ours here has
previously been found~\cite{TothAcin}.

It is interesting to find whether the state has an underlying local realistic model \cite{EPRBell2}. 
Since the classical correlations of the state \(\rho\) are vanishing, we cannot apply the existing multiparty Bell 
inequalities \cite{zukowski1, zukowski2, zukowski3, zukowski4} to test for violation of local realism, as they are based on genuine \(n\)-partite classical correlations. 
For example, the necessary and sufficient conditions, for existence of underlying local realistic models, in 
Ref. \cite{zukowski1, zukowski2, zukowski3, zukowski4}, can only predict that \(\rho\) has a local realistic description for certain numbers of measurement settings 
of the observers. 
The existence of 
such multiparty states requires the derivation of multiparty Bell inequalities that are based on probabilities, {\`a} la Clauser-Horne
inequalities \cite{eita-CH}, or on the concepts of mutual information, entropy, etc. 
If it happens that \(\rho\) does have a local realistic model, then we will have 
a curious scenario, of a ``classical state'' (in the sense that 
it has a local realistic description
even with general tests of local realism that not necessarily involves correlators (cf. \cite{Cerf-paper-PRL})) 
which has no genuine classical correlations.

%It is an interesting problem if the state $\rho$ violates local realism (LR). The standard
%tools to answer this question are the so-called Bell inequalities. However, most of them \cite{WWZB, Werner}
%have the following form $f(t_{A_1\dots A_n}) \leq 0$, where 
%$t_{A_1\dots A_n}=Tr(\rho \sigma_{A_1}\dots\sigma_{A_n})$ and $f$ is some linear function of $t_{A_1\dots A_n}$. 
%Obviously, Bell inequalities of this type cannot 
%be violated by our state.

Another
approach 
%However, there is an alternative 
%way 
to address the problem of local realism (at least numerically), as presented in the Ref. 
\cite{Kaszlikowski}, is to
%  In this approach, one can 
numerically check if the full set of \emph{probabilities} involved 
%a given quantum state admits a local realistic
%description for 
in 
an experiment on the \(n\)-particle state \(\rho\), where the $k$-th observer ($k=1,2,\dots, n$) chooses from a set of $m_k$ 
von Neumann measurements on their qubit, admits a local realistic description. 
This powerful method gives a sufficient and necessary conditions for
the existence of a local realistic  description.
We have performed numerical simulations for the state $\rho$, for three and five qubits, using the above approach.
The simulation was made under the assumption that each observer chooses up to four measurement settings on their
qubit. In all the cases, a local realistic description exists. Interestingly, 
the state $\rho_{\epsilon}=(\frac{1}{2}+\epsilon)|W\rangle\langle W|+
(\frac{1}{2}-\epsilon)|\bar{W}\rangle\langle\bar{W}|$ violates local realism for $\epsilon > 10^{-3}$ (numerical precision used in
the simulation was $10^{-5}$), indicating that even a small perturbation to the state $\rho$, drastically changes its 
behavior, from the point of view local realistic theories. Note that for any $\epsilon >0$ the state $\rho_{\epsilon}$ has genuine 
$n$-partite classical correlations.

It is important to mention here that multipartite \(W\)-states have been prepared in the laboratory by several groups in 
different systems (see e.g. \cite{E1,E2}). Therefore, it is reasonable to hope that 
the effect discussed in this paper can be seen in the laboratory, especially because the effect can already be seen 
for three qubits. 
A possible way to experimentally prepare the state \(\rho\) is by preparing an \(n+1\) qubit \emph{pure} state 
\(\frac{1}{\sqrt{2}}\left(|a\rangle \otimes \ket{W} + |b\rangle \otimes \ket{\overline{W}}\right)\)
where \(\ket{a}\) and \(\ket{b}\) are orthogonal states of the \(n+1\)th qubit, and subsequently tracing out the \(n+1\)th qubit.
In fact, very recently, this four qubit pure state for \(n=3\) has been experimentally prepared by using spontaneous 
parametric down converted photons \cite{E3}.

{\bf Conclusion.}
We have proposed a plausible notion of genuine $n$-party classical
correlations in a multipartite quantum state, and demonstrated
by an explicit example that it is possible for a state to be genuinely
$n$-party entangled without it having genuine $n$-party
classical correlation -- at least for odd $n$ (we expect the same
to occur for even $n$ but don't have an example as yet).
% at the moment).

One possible reaction to this is to dispute the soundness of
our definition of classical correlations -- however, it is not at
all straightforward to come up with a reasonable notion instead
(e.g., there is no ready-to-use entropic correlation measure for $n$-party
systems, otherwise we could try to follow in the steps of~\cite{mutual-I}),
apart from insisting that quantum correlations should imply
classical correlations. One should be cautious, however, with
this intuition, too: in~\cite{random} examples of asymptotically
large bipartite states are presented with almost no mutual information
(i.e., classical correlation) compared to almost maximal entanglement
of formation. See also the recent preprint \cite{PS-R1}
(cf. \cite{PS-R2}), 
where
a different way of separating quantum from classical correlations
is explored, with similar results.

Hence, for the moment, we have to be prepared to accept the paradoxical
statement that quantum correlations can exist without accompanying
classical correlations. What are the consequences? One is to the
Bell's inequalities, most of which -- as already remarked -- are
expressed in terms of $n$-party correlators.
For bipartite systems, there already exist Bell inequalities 
that use statistics beyond correlators \cite{eita-CH, Dago-Gisin-Cerf1, Dago-Gisin-Cerf2, Dago-Gisin-Cerf3}!
However, in multipartite systems, such inequalities are absent.
The  example provided in this paper leads to the need for new Bell inequalities,
based on concepts different than classical correlations
to detect multiparticle nonclassicality.

The boundary between the classical and the quantum worlds is a long-standing and 
arguably hard problem in the foundations of physics. There are many ways of looking at 
this boundary. We believe that the existence of a state that has quantum correlations and yet has 
vanishing classical correlators is a splendid example to deal with the question, e.g., by using 
``local realism'' to characterise the classical 
world.

Along with its direct effects in the science of quantum information, 
the fact can have important consequences in the physics 
of phase transitions \cite{Sachdev}, where so far the usual method 
to detect a phase transition is to look at the scaling of classical 
and quantum correlations in the system;
see~\cite{mutual-I} for an indication that most recently
condensed matter physicists are abandoning correlators in favour
of universal, entropic, measures of correlation. These tools,
however, are restricted to bipartite correlations.
The existence of states with vanishing classical correlations but with  
non-vanishing quantum correlations opens up the possibility of 
phase transitions that are detectable by quantum correlations only.

Finally let us mention here that 
important examples exist where two-body correlations are not enough to describe the 
important phases/properties of the system. And then researchers have resorted to many-body parameters. 
This, for instance, is the case in the Affleck-Kennedy-Lieb-Tasaki system, where a certain ``string order'' 
is necessary (see e.g. \cite{AKLT-string}, and references therein).  
The concept of ``localizable entanglement'' from quantum information, has been a very successful one in 
describing many-body systems, and again it depends on all the particles of the system \cite{localizable-ent}.
Note also that the usual intractability of many-body parameters of condensed matter systems
may change in foreseeable future, with the advent of experimentally realizable quantum simulators: 
Many-body quantum correlated states are being realized in several physical systems, ranging from ion traps to down-converted 
photons in several laboratories around the globe.

AS and US thank the National University of Singapore for their hospitality.
We thank J. Eisert, A. Grudka, O. G{\" u}hne, M. Horodecki, P. Horodecki, 
R. Horodecki, M. Piani, and M. Plenio for comments.
We acknowledge support from the EU IPs SCALA and QAP, ESF QUDEDIS,
%Deutsche Forschungsgemeinschaft
%(SFB 407, SPP 1078, SPP 1116), 
 Spanish MEC (FIS-2005-04627, Consolider Project
QOIT, Acciones Integradas, \& Ram{\'o}n y Cajal),  QIT strategic grant R-144-000-190-646,
U.K.~Engineering and Physical Sciences Research Council
and the Royal Society of the U.K.

\end{document}